\documentclass[twocolumn,aps,prl,showpacs]{revtex4}

\usepackage{latexsym,amssymb,float}
\usepackage{setspace}
\usepackage{graphicx}
\usepackage{epsfig}
\usepackage{amsmath}
\usepackage{bm}

\begin{document}

\title{Differential Conductance and Quantum Interference in Kondo Systems}
\author{Jeremy Figgins and Dirk K. Morr}
\affiliation{Department of Physics, University of Illinois at
Chicago, Chicago, IL 60607, USA}
\date{\today}

\begin{abstract}

We present a large-$N$ theory for the differential conductance,
$dI/dV$, in Kondo systems measured via scanning tunneling
spectroscopy. We demonstrate that quantum interference between
tunneling processes into the conduction band and into the magnetic
$f$-electron states is crucial in determining the experimental {\it
Fano} lineshape of $dI/dV$. This allows one to uniquely extract the
Kondo coupling and the ratio of the tunneling amplitudes from the
experimental $dI/dV$ curve. Finally, we show that $dI/dV$ directly
reflects the strength of the antiferromagnetic interaction in Kondo
lattice systems.

\end{abstract}


\maketitle

Recent progress in scanning tunneling spectroscopy (STS) techniques
have made it possible for the first time to measure the differential
conductance, $dI/dV$, in heavy-fermion compounds \cite{Schm09}.
These materials, whose essential ingredients are a (Kondo) lattice
of magnetic moments that is coupled to a conduction band
\cite{Don77}, exhibits a variety of puzzling phenomena, ranging from
non-Fermi-liquid behavior to unconventional superconductivity
\cite{Exp}. Their microscopic origin likely lies in the competition
between an antiferromagnetic ordering of the magnetic moments, and
their screening by conduction electrons \cite{Don77}, though no
theoretical consensus has emerged as yet \cite{Theory}. STS
experiments, by providing insight into the local electronic
structure \cite{Schm09} of heavy-fermion materials, might hold the
key to understanding their complex properties. The theoretical
challenge in the interpretation of the differential conductance in
Kondo lattice systems \cite{Mal09}, and around single Kondo
impurities \cite{Mad98,Li98,Ujs00,Mad01,Man00} arises from the
quantum interference between electrons tunneling into the conduction
band and into the magnetic $f$-electron states. While $dI/dV$ for a
single Kondo impurity has been successfully described
\cite{Mad98,Li98,Ujs00,Mad01} using a phenomenological form derived
by Fano \cite{Fano61}, a microscopic understanding of how the
interplay between the strength of the Kondo coupling, the
interaction between the magnetic moments, the electronic structure
of the screening conduction band, and quantum interference
determines the $dI/dV$ lineshape, is still lacking.

In this Letter, we address this issue within the framework of a
large-$N$ theory and identify the microscopic origin for the form of
the differential conductance not only around single Kondo impurities
but also in Kondo lattice systems. In particular, we demonstrate
that the lineshape as well as the spatial dependence of $dI/dV$
sensitively depend on the particle-hole asymmetry of the (screening)
conduction band, as well as the quantum interference between the two
tunneling paths. For a single Kondo impurity, this sensitivity
allows one to uniquely extract the Kondo coupling, $J$, as well as
ratio of the tunneling amplitudes into the conduction band and
magnetic $f$-electron states, $t_c$ and $t_f$, respectively, from
the experimental STS data. In addition, for a Kondo lattice, the
$dI/dV$ lineshape provides insight into the strength of the
interaction between the magnetic moments. Due to quantum
interference effects, which can lead to a reversal in the asymmetry
of the $dI/dV$ lineshape already for small changes in $t_f/t_c$, the
differential conductance is in general qualitatively different from
the local density of states (LDOS) of either the conduction band or
the $f$-electron states. However, once the pertinent parameters are
extracted from a theoretical fit, we can predict the frequency and
spatial dependence of the LDOS for both bands, as well as the
electronic correlations between them, thus providing important
insight into the complex electronic structure of Kondo systems.

\begin{figure}[!h]
\includegraphics[width=6.5cm]{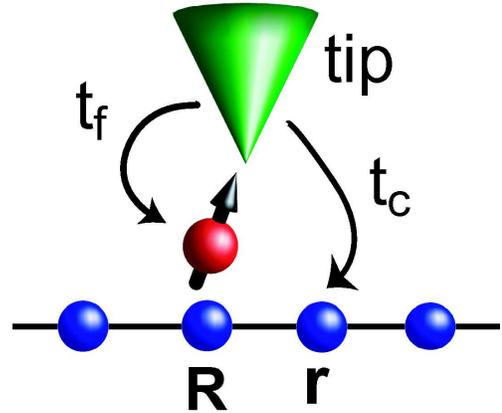}
\caption{(color online) Tunneling paths from the STS tip into the
conduction electron and $f$-electron states with tunneling
amplitudes $t_c$ and $t_f$, respectively.} \label{fig:Fig1}
\end{figure}
We start by considering the differential conductance in a system
with a single Kondo impurity, whose Hamiltonian is given by
\begin{equation}
{\cal H} = -\sum_{{\bf r},{\bf r'},\sigma} t_{{\bf r}{\bf r'}}
c^\dagger_{{\bf r},\sigma} c_{{\bf r}',\sigma} + J {\bf S}^{K}_{\bf
R} \cdot {\bf s}^c_{\bf R} \ , \label{eq:1}
\end{equation}
where $t_{{\bf r}{\bf r'}}$ is the fermionic hopping element between
sites ${\bf r}$ and ${\bf r'}$ of the conduction band,
$c^\dagger_{{\bf r},\sigma}$ ($c_{{\bf r},\sigma}$) creates
(annihilates) a fermion with spin $\sigma$ at site ${\bf r}$, and
the sums run over all sites of the conduction band. ${\bf
S}^{K}_{\bf R}$ and ${\bf s}^c_{\bf R}$ are the spin operators of
the Kondo impurity and the conduction electrons at site ${\bf R}$,
respectively, and $J>0$ is the Kondo coupling. To describe the Kondo
screening of the magnetic impurity, we employ a {\em large-N}
expansion \cite{Coq69,Read83,Col83,Bic87,Mil87,Sen04,Ros06} in which
${\bf S}^{K}_{\bf R}$ is generalized to $SU(N)$ and represented via
Abrikosov pseudofermions $f^\dagger_m, f_m$. These obey the
constraint $\sum_{m=1..N} f^\dagger_m f_m=1$ where $N=2S+1$ is the
spin degeneracy of the magnetic impurity. This constraint is
enforced by means of a Lagrange multiplier $\varepsilon_f$, while
the exchange interaction in Eq.(\ref{eq:1}) is decoupled via a
hybridization field, $s$. On the saddle point level, $\varepsilon_f$
and $s^2$ are obtained by minimizing the effective action
\cite{Read83}. Finally, the tunneling process into a conduction
electron state at ${\bf r}$ and the $f$-electron state at ${\bf R}$,
as schematically shown in Fig.~\ref{fig:Fig1}, is described by
\begin{equation}
{\cal H}_T = \sum_\sigma t_c c^\dagger_{{\bf r},\sigma} d_{\sigma} +
t_f f^\dagger_{{\bf R},\sigma} d_{\sigma} + H.c. \ , \label{eq:tun}
\end{equation}
where $d_{\sigma}$ destroys a fermion in the STS tip. The total
current flowing from the tip into the systems is \cite{Car71}
\begin{eqnarray}
I(V)&=&-\frac{e}{\hbar} \, {\rm Re} \, \int_0^{V} \frac{d \omega}{2
\pi} \left[ t_c \, {\hat G}^K_{12}(\omega) + t_f \, {\hat
G}_{13}^K(\omega)  \right] .
 \label{eq:IV}
\end{eqnarray}
Here
\begin{equation}
{\hat G}^K(\omega) = [{\hat 1} - {\hat g}^r(\omega) {\hat t}]^{-1}
{\hat f}_0(\omega) [{\hat 1} -  {\hat t} {\hat g}^a(\omega) ]^{-1}
\end{equation}
is the full Keldysh Greens function matrix, ${\hat t} $ is the
symmetric hopping matrix with ${\hat t}_{12} = t_c$, ${\hat t}_{13}
= t_f$, and zero otherwise. ${\hat g}^r(\omega)$ and ${\hat
f}_0(\omega)$ are the retarded and Keldysh Greens function matrices
of the Kondo system with
\begin{eqnarray}
{\hat f}_0(\omega) &=& 2i \left(1-2 {\hat n_F}(\omega) \right) {\rm
Im} \left[ {\hat g}^r(\omega) \right] \ ; \nonumber \\
{\hat g}^r(\omega) &=&
\begin{pmatrix} g^r_t(\omega) & 0 & 0 \\ 0 & g^r_{cc}({\bf r}, {\bf
r}, \omega)& g^r_{cf}({\bf r}, {\bf R},
\omega) \\
0 & g^r_{fc}({\bf R}, {\bf r}, \omega) & g^r_{ff}({\bf R}, {\bf R},
\omega)
\end{pmatrix} \ ,
\end{eqnarray}
where $g^r_t, g^r_{cc}$ and $g^r_{ff}$ are the local Greens
functions of the tip, conduction and $f$-electron states,
respectively, and $g_{fc}({\bf R}, {\bf r}, \tau)=-\langle T_\tau
f^\dagger_{\bf R}(\tau) c_{\bf r}(0) \rangle$. Moreover,
\begin{equation}
{\hat n_F}(\omega) = \begin{pmatrix} n_F^t(\omega) & 0 & 0 \\
0 & n_F(\omega) & 0 \\ 0 & 0 & n_F(\omega)
\end{pmatrix}
\end{equation}
with $n_F^t$ ($n_F$) being the Fermi-distribution function of the
tip ($f$- and $c$-electron states). While the results shown below
are obtained from Eq.(\ref{eq:IV}) via differentiation, it is
instructive to consider the leading order contributions to $dI/dV$
in the weak-tunneling limit, $t_c,t_f \rightarrow 0$, given by
\begin{eqnarray}
\frac{dI(V)}{dV} & =& \frac{2 \pi e}{\hbar} N_t \left[t_c^2
N_{c}({\bf
r}, V) + t_f^2 N_{f}({\bf R}, V) \right. \nonumber \\
& & \quad  + \left. 2 t_c t_f N_{cf}({\bf r}, {\bf R}, V) \right]
\label{eq:dIdV}
\end{eqnarray}
with $N_t, N_c$ and $N_f$ being the density of states of the tip,
conduction and $f$-electron states, respectively, and $N_{cf}=-{\rm
Im} g^r_{cf}/\pi$. The last term in Eq.(\ref{eq:dIdV}) describes
quantum interference processes between the two tunneling paths,
which, as we show below, are crucial in determining the lineshape of
the differential conductance.

\begin{figure}[h]
\includegraphics[width=8.5cm]{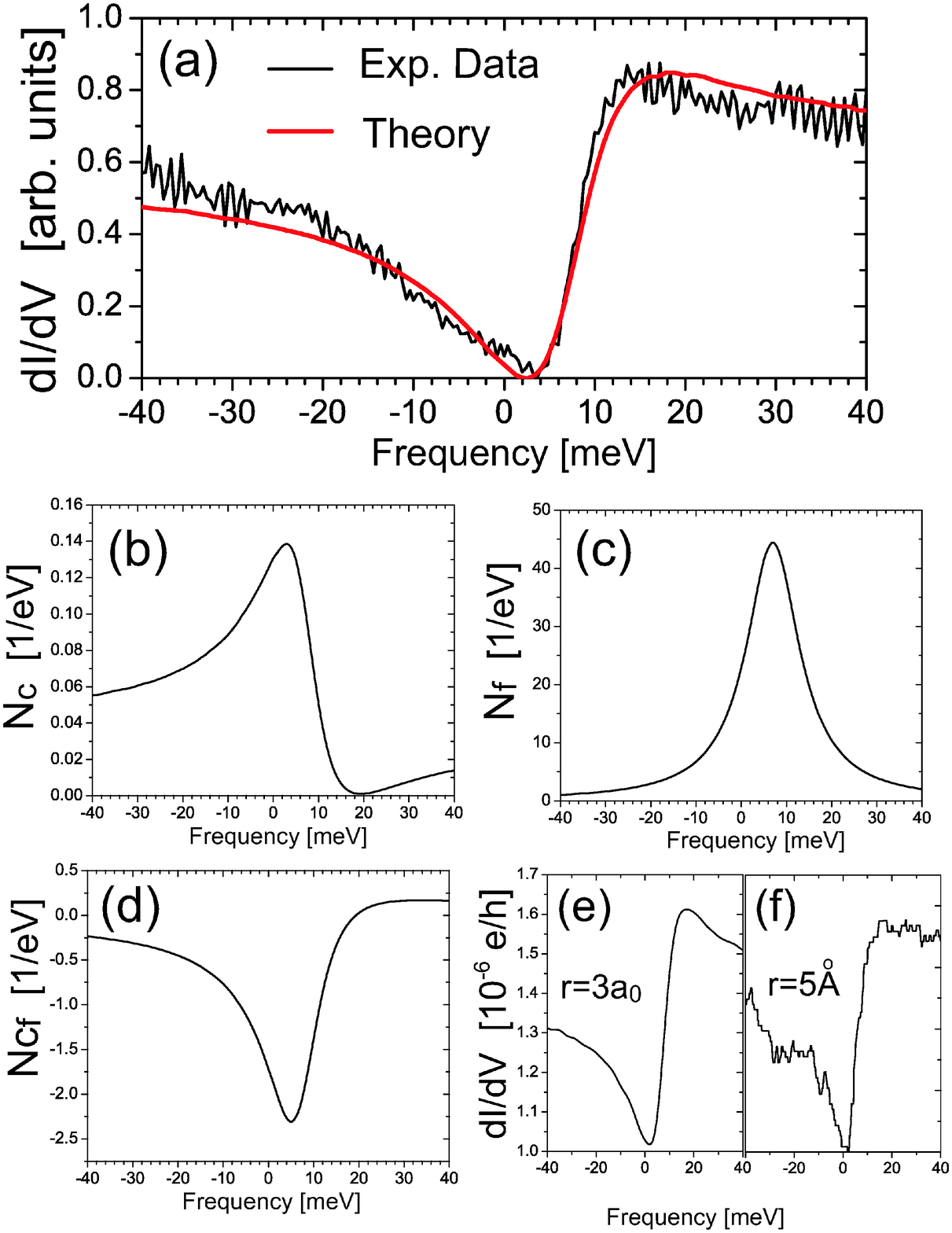}
\caption{(color online) (a) Experimental $dI/dV$ curve of
Ref.~\cite{Mad01} at the site of a Co atom on a Au(111) surface
together with a theoretical fit from Eq.(\ref{eq:IV}) with $N=4$,
$t_f/t_c=0.066$, $t_c=1$ meV, $J=1.39$ eV, $s=250$ meV,
$\varepsilon_f=19$ meV and $N_t=1/eV$. A constant background was
subtracted from the experimental data. (b) conduction electron LDOS,
$N_c(\omega)$ (c) $f$-electron LDOS $N_f(\omega)$, and (d)
$N_{cf}(\omega)$ at ${\bf R}$. (e) $dI/dV$ at a distance of $r=3
a_0$ from the Co atom for $t_f=0$. Parameters in (b)-(e) are the
same as in (a). (f) Experimental $dI/dV$ curve of Ref.~\cite{Mad98}
at $r=5 \AA$ from the Co atom.} \label{fig:ExpFit}
\end{figure}
In Fig.~\ref{fig:ExpFit}(a) we present the experimental $dI/dV$ data
of Ref.~\cite{Mad01} for a tip positioned above a Co atom on a
Au(111) surface together with a theoretical fit obtained from
Eq.(\ref{eq:IV}). Here, tunneling into the conduction band involves
only the state at ${\bf R}$, i.e., ${\bf r}={\bf R}$. The peak and
dip in $dI/dV$ are a direct signature of the hybridization between
the conduction band and the $f$-electron state of the Kondo impurity
and are commonly referred to as the Kondo resonance. As input
parameters, we took the screening conduction band to be given by the
Au(111) surface states possessing a triangular lattice structure
with $t=1.3$ eV and $\mu=-7.34$ eV \cite{Sch09}, and used $N=4$ to
describe the $S=3/2$ Co spin. The theoretical $dI/dV$ curve is then
solely determined by $J$ and $t_f/t_c$, which control the width of
the dip and the overall asymmetry of $dI/dV$, respectively. By
performing an extensive survey, we found that there exists a unique
set of parameters, $J=1.39$ eV and $t_f/t_c=0.0066$, that yield the
good quantitative agreement between the theoretical and experimental
data shown in Fig.~\ref{fig:ExpFit}(a). We note that while the STS
tip is positioned above the Co atom, $t_f/t_c$ is small, likely
reflecting the suppression of the tunneling process into the
$f$-electron state by Coulomb effects. Moreover, once $J$ is
obtained from the fit, we can compute the LDOS of the conduction and
$f$-electron states, which are presented in
Figs.~\ref{fig:ExpFit}(b) and (c), respectively, as well as the
electronic correlations between the two bands, as reflected by
$N_{cf}$ shown in Fig.~\ref{fig:ExpFit}(d). The lineshape of either
LDOS (or of their superposition) is qualitatively different from
that of $dI/dV$, demonstrating the importance of quantum
interference in determining the latter. Finally, as the STS tip is
moved away from the Co atom, direct tunneling into the $f$-electron
state becomes suppressed and $t_f \rightarrow 0$. Therefore, in
Fig.~\ref{fig:ExpFit}(e), we present the theoretical $dI/dV$ curve
with $t_f=0$ at a distance of $r=3 a_0$ from the Co atom. We note
that while $t_f=0$ the asymmetry of $dI/dV$ is now the same as that
at the site of the Co atom, and qualitatively agrees with the
experimental $dI/dV$ curve at $r=5 \AA$ \cite{Mad98} shown in
Fig.~\ref{fig:ExpFit}(f), demonstrating the consistency of our
approach. A more quantitative fit would require an extensive spatial
survey of $dI/dV$ away from the Co atom.

\begin{figure}[!t]
\includegraphics[width=8.5cm]{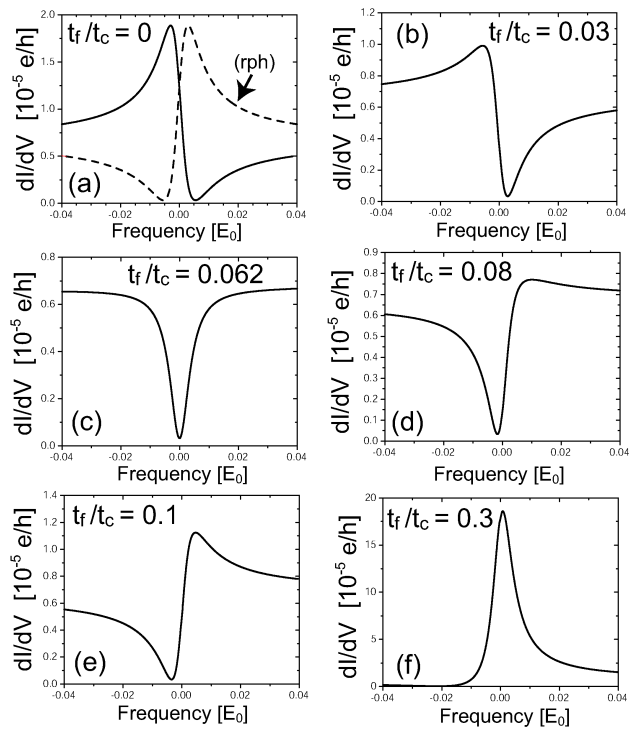}
\caption{(a) - (f) $dI/dV$ at ${\bf r}={\bf R}$ as a function of
energy for $N=2$, $J=0.5 E_0$, $N_t=1.0/E_0$, $t_c=0.001 E_0$,
$\varepsilon_f=0.00520 E_0$, $s=0.0847 E_0$ and different values of
$t_f/t_c$. Dashed line in (a) represents $dI/dV$ for a conduction
band with a reversed particle-hole (rph) asymmetry.}
\label{fig:Fig2}
\end{figure}

The asymmetry of the $dI/dV$ lineshape is determined by two
microscopic properties: the particle-hole asymmetry of the screening
conduction band, and the ratio of the tunneling amplitudes,
$t_f/t_c$. To demonstrate this dependence, we present in
Fig.~\ref{fig:Fig2} the evolution of $dI/dV$ with increasing ratio
$t_f/t_c$. To contrast and complement the results shown in
Fig.~\ref{fig:ExpFit}, we take $N=2$, corresponding to a spin-$1/2$
moment, and consider a conduction band on a square lattice with
$t=0.5 E_0$ and $\mu=-1.809 E_0$. The resulting circular Fermi
surface with Fermi wavelength $\lambda_F = 10 a_0$ is representative
of the Au(111) and Cu(111) surfaces states employed in
Refs.\cite{Mad98,Man00,Mad01}.
For $t_f=0$ [solid line in Fig.~\ref{fig:Fig2}(a)], $dI/dV$ exhibits
a Kondo resonance whose asymmetry is opposite to the experimentally
observed one shown in Fig.~\ref{fig:Fig1}(a). The asymmetry of
$dI/dV$ for $t_f=0$ is a direct consequence of the particle-hole
asymmetry of the conduction band. Indeed, reversing the latter via
$\mu \rightarrow -\mu$, also leads to a reversal of the asymmetry in
$dI/dV$, as shown by the dashed line in Fig.~\ref{fig:Fig2}(a).
Moreover, with increasing $t_f/t_c$, the height of the peak on the
negative energy side, as well as the width of the dip in $dI/dV$
decrease while its minimum shifts to lower energies [see
Fig.~\ref{fig:Fig2}(b)], leading to an almost symmetric $dI/dV$
curve for $t_f/t_c = 0.062$ [see Fig.~\ref{fig:Fig2}(c)]. Increasing
$t_f/t_c$ even further [see Fig.~\ref{fig:Fig2}(d)] now reverses the
asymmetry in $dI/dV$, yielding a peak on the positive energy side,
and a minimum at slightly negative energies. The asymmetry of the
$dI/dV$ lineshape is now similar to that observed experimentally.
However, in contrast to the case of a spin-$3/2$ moment ($N=4$)
considered in Fig.~\ref{fig:ExpFit}(a), the minimum in $dI/dV$ is
located at negative energies for a spin-$1/2$ moment ($N=2$).
Indeed, for $N=2$, no fit to the experimental data of
Fig.~\ref{fig:ExpFit}(a) can be obtained. This demonstrates that the
differential conductance directly reflects the spin of the screened
magnetic moment. Increasing $t_f/t_c$ even further [see
Fig.~\ref{fig:Fig2}(f)] leads to an increase in the height of the
peak and a widening of the dip.

We next turn to the discussion of the differential conductance in a
Kondo lattice system, whose complex properties are determined by the
competition between the Kondo screening of the magnetic moments and
their antiferromagnetic ordering \cite{Don77}. Its Hamiltonian is
obtained by appropriately extending Eq.(\ref{eq:1}), and adding the
term ${\cal H}_I=\sum_{{\bf r,r^\prime} }I_{{\bf r,r^\prime}} {\bf
S}^{K}_{\bf r} {\bf S}^{K}_{\bf r^\prime}$ representing the
antiferromagnetic interaction between the moments. Here, we take
$I_{{\bf r,r^\prime}}>0$ to be non-zero for nearest-neighbor sites
only. Using again an Abrikosov pseudo-fermion representation of
${\bf S}^{K}_{\bf r}$, the Hamiltonian is decoupled by introducing
the spatially uniform mean-fields \cite{Sen04} $s= J \langle
f^\dagger_{{\bf r},\alpha} c_{{\bf r},\alpha} \rangle $ and $\chi_0=
I \langle f^\dagger_{{\bf r},\alpha} f_{{\bf r'},\alpha} \rangle$,
where the latter is a measure of the magnetic correlations in the
system. The constraint $\langle n_f \rangle = 1$ is enforced via a
local on-site energy, $\sum_{\bf r} \varepsilon_f f^\dagger_{{\bf
r},\alpha} f_{{\bf r},\alpha}$.

\begin{figure}[!t]
\includegraphics[width=8.5cm]{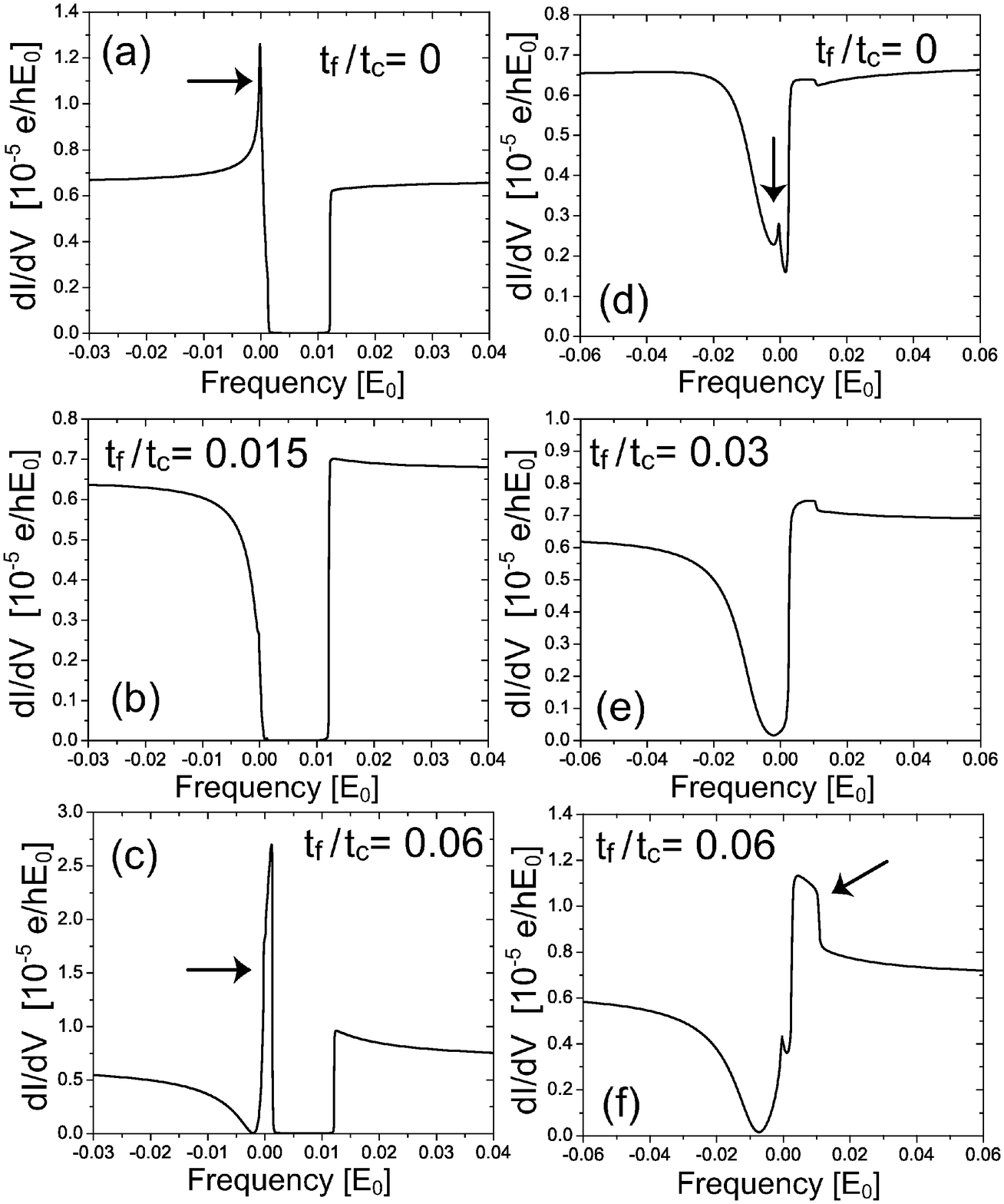}
\caption{Evolution of $dI/dV$ in a Kondo lattice with $t_f/t_c$ for
$N=2$, $J=0.5 E_0$, $N_t=1/E_0$, $t_c=0.001 E_0$, and (a) -  (c)
$I/J=0.001$ with $\varepsilon_f=0.0012 E_0$, $s=0.0485 E_0$,
$\chi_0=0.00017 E_0$, and (d) -  (f) $I/J=0.015$ with
$\varepsilon_f=0.00094 E_0$, $s=0.0480 E_0$ and $\chi_0=0.00259
E_0$.} \label{fig:Fig4}
\end{figure}

The magnetic interactions in the (screened) Kondo lattice have a
profound effect on the form of the differential conductance, as
shown in Fig.~\ref{fig:Fig4} where we present the evolution of
$dI/dV$ with $t_f/t_c$ for two different magnetic interaction
strengths, $I/J=0.001$ (left column) and $I/J=0.015$ (right column).
While for $I/J=0.001$, $dI/dV$ exhibits a hard {\it hybridization
gap} for all values of $t_f/t_c$, only a suppression of the
differential conductance around the Fermi energy is found for
$I/J=0.015$. However, in both cases, $dI/dV$ exhibits a peak on the
negative energy side (indicated by arrows), which arises from the
Van Hove singularity of the large (hybridized) Fermi surface. This
peak is first suppressed with increasing $t_f/t_c$ [see
Figs.~\ref{fig:Fig4}(b) and (e)], but then reemerges together with a
second peak [indicated by arrows in Figs.~\ref{fig:Fig4}(c) and
(f)], which is the precursor of the emerging $f$-electron band. This
second peak is centered around the Fermi energy for $I/J=0.001$, but
is located at positive energies for $I/J=0.015$. In the latter case,
we also find a shift of the minimum in $dI/dV$ to lower energies
with increasing $t_f/t_c$. This qualitative difference in the
differential conductance thus provides direct insight into the
strength of the antiferromagnetic interactions.

In summary, we have presented a large-$N$ theory for the
differential conductance in Kondo systems. We demonstrated that
quantum interference between tunneling paths is crucial in
explaining the experimentally observed {\it Fano} lineshape of
$dI/dV$. This allows one to uniquely extract the Kondo coupling as
well as the ratio of the tunneling amplitudes from the experimental
$dI/dV$ curve. Finally, we showed that $dI/dV$ reflects the strength
of the antiferromagnetic interaction in Kondo lattice systems.

We would like to thank J.C. Davis, V. Madhavan, and H. Manoharan for
stimulating discussions, and V. Madhavan for providing the data of
Ref.~\cite{Mad01}. D.K.M. would like to thank the James Franck
Institute at the University of Chicago for its hospitality during
various stages of this project. This work is supported by the U.S.
Department of Energy under Award No. DE-FG02-05ER46225.

\end{document}